\documentclass[%
 reprint,
 superscriptaddress,
 amsmath,amssymb,
 aps,
 prd,
]{revtex4-2}

\usepackage{graphicx}
\usepackage{dcolumn}
\usepackage{bm}
\usepackage{color}

\newcommand{\jcap}{Journal of Cosmology and Astroparticle Physics}
\newcommand{\pasj}{Publications of the Astronomical Society of Japan}
\newcommand{\apjs}{The Astrophysical Journal Supplement Supplement Series}
\newcommand{\nphysa}{Nuclear Physics A}
\newcommand{\apss}{Astrophysics and Space Science}
\newcommand{\apjl}{The Astrophysical Journal Letters}

\begin{document}

\preprint{APS/123-QED}

\title{Comparative Testing of Subgrid Models for Fast Neutrino Flavor
Conversions in Core-collapse Supernova Simulations}

\author{Ryuichiro Akaho}
\affiliation{Faculty of Science and Engineering, Waseda University, 3-4-1 Okubo, Shinjuku, Tokyo 169-8555, Japan}

\author{Hiroki Nagakura}
\affiliation{Division of Science, National Astronomical Observatory of Japan, 2-21-1 Osawa, Mitaka, Tokyo 181-8588, Japan}

\author{Shoichi Yamada}
\affiliation{Advanced Research Institute for Science and Engineering, Waseda University, 3-4-1 Okubo, Shinjuku, Tokyo 169-8555, Japan}

\date{\today}

\begin{abstract}
We investigate key methodologies of Bhatnagar-Gross-Krook subgrid modeling for neutrino fast flavor conversions (FFC) in core-collapse supernova based on spherically symmetric
Boltzmann radiation hydrodynamics simulations.
We first examine time integration methods (explicit, implicit, or semi-implicit) and time step control for the subgrid term, and then compare various approaches in the literature approximating FFCs in two aspects:
(1) angular dependent survival probability of neutrinos versus simple equipartition condition with a certain baryon mass density threshold, and (2) 4-species treatment versus 3-species assumption ($\nu_x=\bar\nu_x$).
We find that the equipartition condition is reasonable for out-going neutrinos, but large deviations emerge in the ingoing neutrinos, that has an influence on matter profiles.
We also find that the 3-species model, in which flavor conversions evolve towards erasing electron neutrino lepton number (ELN) crossings, behave differently from the 4-species models where heavy leptonic neutrino number (XLN) are appropriately treated in FFC subgrid modeling.
In 4-species models, we commonly observe noticeable differences between $\nu_x$ and $\bar\nu_x$, highlighting the limitation in 3-species treatments to study impacts of flavor conversion on neutrino signals.
Our result also suggests that FFC models yield lower neutrino heating rate and smaller shock radii compared to cases with no FFC, in agreement with earlier studies employing quantum kinetic neutrino transport.
This work provides valuable information towards robust implementation of FFC subgrid model into classical transport, and serves as a pilot study for future multi-dimensional simulations.
\end{abstract}

\maketitle


\section{Introduction}

Core-collapse supernovae (CCSNe) mark the death of massive stars ($M\gtrsim8M_\odot$), which result in the birth of compact objects such as neutron stars (NSs) or black holes (BHs)
(see \cite{Janka2016ARNPS..66..341J,Janka2017hsn..book.1095J,Janka2017hsn..book.1575J,Mezzacappa2020LRCA....6....4M,Burrows2021Natur.589...29B,Mezzacappa2023IAUS..362..215M,Boccioli2024Univ...10..148B,Yamada2024PJAB..100..190Y,Suzuki2024PTEP.2024eB101S,Janka2025arXiv250214836J} for recent reviews).
Neutrinos are known to be the main driver of the explosion and a valuable information source to probe the core as well.
Extensive efforts have been made for the realistic modeling of CCSNe, and some recent simulations successfully reproduced explosion energies and nucleosynthetic yields suggested by optical observations \cite{Burrows2021Natur.589...29B,Janka2024Ap&SS.369...80J}.
However, the effect of neutrino oscillations on CCSN dynamics has often been ignored.
In reality, under the dense neutrino environment as CCSNe, collective neutrino oscillation induced by neutrino-neutrino self-interaction is known to occur
(see \cite{Duan2010ARNPS..60..569D,Richers2022arXiv220703561R,Volpe2024RvMP...96b5004V,Johns2025arXiv250305959J} for recent reviews).

Among the instabilities associated with these collective oscillations, fast flavor instability (FFI), which induces fast flavor conversion (FFC) \cite{Sawyer2005PhRvD..72d5003S,Duan2006PhRvD..74j5014D,Duan2006PhRvD..74l3004D,Sawyer2009PhRvD..79j5003S}, is gaining great attention.
The existence of FFI is known to be equivalent to the angular crossings in momentum space \cite{Morinaga2022PhRvD.105j1301M,Dasgupta2022PhRvL.128h1102D}, and the onset of FFC typically acts to eliminate these crossings \cite{Zaizen2023PhRvD.107j3022Z}.
Post-process analyses of FFI in CCSNe simulations have been performed 
\cite{Nagakura2019ApJ...886..139N,DelfanAzari2019PhRvD..99j3011D,Abbar2020PhRvD.101d3016A,DelfanAzari2020PhRvD.101b3018D,Glas2020PhRvD.101f3001G,Nagakura2021PhRvD.104h3025N,Harada2022ApJ...924..109H,Akaho2023ApJ...944...60A,Akaho2024PhRvD.109b3012A,Xiong2025arXiv250519592X}
and it is now widely believed that it ubiquitously appears in CCSNe.

Although the occurrence of FFC and its nonlinear behavior has been extensively studied by various methods including linear stability analysis and numerical simulations
\cite{Morinaga2018PhRvD..97b3024M,Morinaga2020PhRvR...2a2046M,Bhattacharyya2020PhRvD.102f3018B,Morinaga2020PhRvR...2a3045M,Morinaga2021PhRvD.103h3014M,Bhattacharyya2021PhRvL.126f1302B,Dasgupta2022PhRvL.128h1102D,Xiong2023PhRvD.108f3003X,Zaizen2023PhRvD.107l3021Z,Fiorillo2024JHEP...08..225F,Fiorillo2024arXiv240917232F,Fiorillo2024arXiv241209027F,Fiorillo2024PhRvL.133v1004F}, its effect on CCSN dynamics is still unclear.
The difficulty lies in solving the quantum kinetic equation (QKE)
\cite{Sigl1993NuPhB.406..423S,McKellar1994PhRvD..49.2710M,Serreau2014PhRvD..90l5040S,Vlasenko2014PhRvD..89j5004V,Kartavtsev2015PhRvD..91l5020K,Blaschke2016PhRvD..94c3009B,Froustey2020JCAP...12..015F,Kainulainen2024JHEP...02..217K,Kainulainen2024-dq}, which is required to describe collective neutrino oscillations.
The characteristic length/time-scale of collective neutrino oscillation can be orders of magnitude shorter than those typically used in classical neutrino transport simulations.
In addition, lack of momentum space angle is known to cause the appearance of the spurious modes \cite{Sarikas2012PhRvD..86l5020S}, hence QKE simulations typically requires much finer angle resolution in momentum space than classical Boltzmann transport, at least $\sim100$ mesh points \cite{Zaizen2023PhRvD.107l3021Z,Nagakura2023PhRvL.130u1401N,Xiong2024PhRvD.109l3008X}.
Direct simulation of QKE with CCSN background has been performed recently \cite{Shalgar2023PhRvD.107l3004S,Shalgar2023PhRvD.107f3025S,Nagakura2023PhRvL.130u1401N,Nagakura2022PhRvL.129z1101N,Nagakura2023PhRvD.108l3003N,Nagakura2023PhRvD.108j3014N,Nagakura2023PhRvL.130u1401N,Xiong2024PhRvD.109l3008X}, but often assume spherical symmetry and limited to simulation duration much shorter than the CCSN explosion timescale ($\sim100\,\mathrm{ms}$).

Given that direct simulation of QKE is unrealistic to simulate CCSN explosion timescale, an alternative approach is taken to investigate the effect of FFC on CCSNe; to phenomenologically include FFC effects on classical simulations
\cite{Ehring2023PhRvD.107j3034E,Ehring2023PhRvL.131f1401E,Mori2025PASJ..tmp...18M,Wang2025arXiv250304896W}
(see also \cite{Just2022PhRvD.105h3024J,Qiu2025arXiv250311758Q} for application to neutron star merger simulations).
Approximate transport methods are typically used for CCSN simulations, which does not provide enough information to detect flavor instability. 
Hence simplified FFC subgrid models are usually used.
Reflecting the nature of FFC that occurs in the semi-transparent to optically thin region, forcing flavor equipartition below a certain density threshold is sometimes used \cite{Ehring2023PhRvD.107j3034E,Ehring2023PhRvL.131f1401E,Mori2025PASJ..tmp...18M}.
More realistic approach is to reconstruct momentum space distribution assuming maximum entropy distribution \cite{Wang2025arXiv250304896W} to detect angular crossings.
However, the angular distribution provided by the maximum entropy assumption deviates from Boltzmann transport (see \cite{Iwakami2022ApJ...933...91I} for the comparison of closure relation).
The self-consistent subgrid modeling of FFC requires multi-angle Boltzmann transport, which is already successful under fixed hydrodynamics background \cite{Nagakura2024PhRvD.109h3013N,Xiong2025PhRvL.134e1003X} (also see attempts on moment transport using machine learning \cite{Abbar2023PhRvD.107j3006A,Abbar2024PhRvD.109b3033A,Abbar2024PhRvD.109h3019A,Abbar2024PhRvD.109d3024A}).

Another problematic assumptions typically made in previous studies is the
3-species assumption, regarding heavy-lepton type neutrino and antineutrinos to be identical ($\nu_x=\bar\nu_x$) \cite{Ehring2023PhRvD.107j3034E,Ehring2023PhRvL.131f1401E,Qiu2025arXiv250311758Q,Wang2025arXiv250304896W,Mori2025PASJ..tmp...18M}.
With this assumption, unphysical mixing between $\nu_e$ {(electron-type neutrino)} and $\bar\nu_e$ {(electron-type anti-neutrino)} occur through $\nu_x$.
In addition, the instability condition changes; in 4-species case, FFC is driven by ELN-XLN crossings (differences between electron and heavy-lepton types), whereas in 3-species case, FFC is driven solely from ELN crossings.
The occurrence of FFC changes when the difference between $\nu_x$ and $\bar\nu_x$ becomes prominent, such as the distribution after the flavor conversion.

We address aforementioned issues through Boltzmann neutrino radiation hydrodynamics simulation coupled with Bhatnagar-Gross-Krook (BGK) \cite{Bhatnagar1954PhRv...94..511B} subgrid modeling of FFC, with 4-species treatment.
The occurrence of FFC and the asymptotic distribution can be self-consistently treated because full momentum space distribution is treated in the Boltzmann transport.
We solve for 4-species ($\nu_e$, $\bar\nu_e$, $\nu_x$, $\bar\nu_x$), and also perform simulations with 3-species assumption to compare its effect.
We also perform simple equipartition mixing approach with the density threshold and compare the results with the angle-dependent subgrid modeling.
This is meant for a pilot study to compare subgrid modeling methods, and prepare for a multi-dimensional simulations in the future. 

This paper is organized as follows.
Numerical details are explained in Sec. \ref{sec:numerical}.
The time discretization methods are described and tested in Sec. \ref{sec:timetest}.
Flavor evolution at the appearance of instability is discussed in Sec. \ref{sec:early}, and effects of FFC on CCSN dynamics is discussed in Sec. \ref{sec:long}. Sec. \ref{sec:summary} concludes our findings.
The natural unit $c=\hbar=1$ is employed, where $c$ and $\hbar$ denote the speed of light and the Planck constant, respectively. The Greek and Latin indices run over 0 to 3, and 1 to 3, respectively.

\section{Numerical Method}
\label{sec:numerical}

\subsection{Boltzmann Neutrino Radiation Hydrodynamics}
\label{sec:boltz}

We use general relativistic Boltzmann neutrino radiation hydrodynamics code \cite{Nagakura2014ApJS..214...16N,Nagakura2017ApJS..229...42N,Nagakura2019ApJS..240...38N,Akaho2021ApJ...909..210A,Akaho2023ApJ...944...60A}.
Although our code is capable of treating spatial multi-dimension, we focus on spherically symmetric simulations in this paper. 
We briefly summarize numerical methods below.
{Code verification testes and the further details are described in the papers cited above.}

Boltzmann equation {with respect to the phase space distribution function $f$} written in the conservative form \cite{Shibata2014PhRvD..89h4073S} limited to spherical symmetry can be written as
\begin{eqnarray}
\label{eq_conservBoltz}
\frac{1}{\alpha}\frac{\partial f}{\partial t} 
&+& \frac{\mathrm{cos}\theta_\nu}{\alpha r^2 \sqrt{\gamma_{rr}}}\frac{\partial}{\partial r}\left(\alpha r^2 f\right)
-\frac{1}{\epsilon^2}\frac{\partial}{\partial\epsilon}\left(\epsilon^3f\omega_{(t)}\right) \nonumber \\
&+& \frac{1}{\mathrm{\mathrm{sin}\theta_\nu}}\frac{\partial}{\partial\theta_\nu}\left(\mathrm{sin}\theta_\nu f\omega_{(\theta_\nu)}\right) = S_{\rm rad} + S_{\rm osc},
\end{eqnarray}
where {$t$, $r$, $\epsilon$, $\theta_\nu$}, $\alpha$, and $\gamma_{ij}$ denote the {time, radius, energy, zenith angle in momentum space,} lapse function and the spatial metric, respectively. 
{Under 1D, the dependences on $\theta$, $\phi$ (zenith and azimuth angles in configuration space) and $\phi_\nu$ (azimuth angle in momentum space) are dropped.}
The left hand side represents the {neutrino transport in the phase space}, and the first term in the right hand side {($S_{\rm rad}$)} represent the collision terms for the classical Boltzmann equation. 
The second term on the right hand side, $S_{\rm osc}$ represents the effects of FFC, which is calculated based on the method described in Sec. \ref{sec:subgrid}.
The factors $\omega_{(t)}$ and $\omega_{(\theta_\nu)}$ are defined as
\begin{eqnarray}
\omega_{(t)} &\equiv& \epsilon^{-2} p^\mu p_\nu \nabla_\mu e_{(t)}^\nu, \nonumber \\
\omega_{(\theta_\nu)} &\equiv& - \epsilon^{-2} p^\mu p_\nu \nabla_\mu e_{(r)}^\nu\mathrm{sin}\theta_\nu, 
\end{eqnarray}
where $p^\mu$ denotes the neutrino four-momentum.
{The covariant derivative is denoted with $\nabla$.}
The tetrad bases are given as
\begin{eqnarray}
e^\mu_{(t)} & \equiv& \left(\alpha^{-1}, 0,0,0\right),
\nonumber \\
e^\mu_{(r)} & \equiv& 
\left(0,\gamma_{rr}^{-1/2},0,0\right), \label{eq_tetrad}
\end{eqnarray}

General relativistic hydrodynamical equation \cite{Shibata2016nure.book.....S} limited to spherical symmetry becomes:
\begin{equation}\label{eq_massconserv}
\partial_t \rho_* + \partial_r(\rho_* v^r) = 0,
\end{equation}
\begin{eqnarray}\label{eq_Euler}
&& \partial_t S_r + \partial_r(S_r v^r + \alpha\sqrt{\gamma}P) 
\nonumber \\
&=& -S_0 \partial_r \alpha - \frac{1}{2}\alpha \sqrt{\gamma} S_{jk}\partial_r \gamma^{jk} - \alpha \sqrt{\gamma}\gamma_r{}^\mu G_\mu, 
\end{eqnarray}
\begin{eqnarray}\label{eq_energyconserv}
&& \partial_t (S_0 - \rho_*) + \partial_r ((S_0 - \rho_*)v^r + \sqrt{\gamma}Pv^k) 
\nonumber \\
&=& \alpha\sqrt{\gamma}S^{ij}K_{ij} - S_r D^r \alpha + \alpha \sqrt{\gamma}n^\mu G_\mu,
\end{eqnarray}
{
\begin{equation}
\label{eq:ye}
\partial_t(\rho_\ast Y_e) + \partial_j (\rho_\ast Y_e v^j) = - \alpha \sqrt{\gamma} \Gamma 
\end{equation}}
where
\begin{eqnarray}
&&v^r \equiv {u^r}/{u^t}, \qquad
\rho_* \equiv \alpha \sqrt{\gamma}\rho_0 u^t = \sqrt{\gamma}\rho_0\omega, \\
&&S_r \equiv \rho_* h u_r c, \qquad
S_0 \equiv\sqrt{\gamma}(\rho h w^2 - P), \\
&& w \equiv \alpha u^t, \qquad
S_{ij} \equiv\rho h u_i u_j + P\gamma_{ij},
\end{eqnarray}
and $\rho$, $P$, $u^\mu$, $h$, {$Y_e$} represent the density, the pressure, the four velocity, specific enthalpy, {and the electron fraction}, respectively. 
{Eqs. \ref{eq_massconserv}, \ref{eq_Euler}, \ref{eq_energyconserv} and \ref{eq:ye} are continuity, Euler, energy conservation and lepton-number conservation equations.}
The symbol $G^\mu$ {and $\Gamma$} stand for the feedback from neutrinos {, where $G^0$ represent the energy feedback, $G^i$'s represent the momentum feedback.}
{
They are related to the collision terms of the Boltzmann equation as
\begin{eqnarray}
&&G^\mu\equiv \sum_i \int p^\mu_i \,S_{\mathrm{rad}(i)}\, \epsilon^2 d\epsilon d(\cos\theta_\nu)d\phi_\nu, \\
&&\Gamma\equiv \int (S_{\mathrm{rad}(\nu_e)}-S_{\mathrm{rad}(\bar\nu_e)})\, \epsilon^2 d\epsilon d(\cos\theta_\nu)d\phi_\nu.
\end{eqnarray}
The subscript $i$ denote the sum over species.
Since we take into account charged-current interactions only for electron-type, the sum is taken only for $\nu_e$ and $\bar\nu_e$ for the calculation of $\Gamma$.
}

Under spherically symmetric condition, we assume the metric ansatz as 
\begin{equation}
\alpha = e^{\Phi(t,r)},\qquad \gamma_{rr} = \left(1-\frac{2m(t,r)}{r}\right)^{-1},
\end{equation}
where there are only two independent functions $m$ and $\Phi$, which only depend on $t$ and $r$.
Functions $m$ and $\Phi$ can be obtained by solving the ordinary differential equations at each time step 
\begin{equation}
\label{eq_dmdr}
\frac{\partial m}{\partial r} = 4 \pi r^2 (\rho h w^2 - P),
\end{equation}
\begin{equation}
\label{eq_dphidr}
\frac{\partial \Phi}{\partial r} = \left(1-\frac{2m}{r}\right)^{-1}
\left(\frac{m}{r^2}+4\pi r(\rho h v^2 + P)\right).
\end{equation}
{In the Newtonian limit, $m$ and $\Phi$ coincides with the enclosed mass and the gravitational potential, respectively.}

As for the neutrino-matter interactions, in addition to the standard set \cite{Bruenn1985ApJS...58..771B}, nucleon-nucleon Bremsstrahlung and the neutrino-electron capture on light and heavy nuclei \citep{Langanke2000NuPhA.673..481L,Langanke2003PhRvL..90x1102L,Juodagalvis2010NuPhA.848..454J} are incorporated.
As for the nuclear matter, Furusawa-Togashi equation of state \cite{Furusawa2017JPhG...44i4001F} based on the variational method is used.

Radial grid covers $r\in\left[0:5000\right]\,\mathrm{km}$ with $384$ mesh points.
The zenith angle grid in momentum space covers $\theta_\nu\in[0:\pi]$ with $10$ mesh points, and the energy grid covers $\epsilon\in\left[0:300\right]\,\mathrm{MeV}$ with 20 mesh points. {The zenith angle is measured with respect to the radially outgoing direction (see Fig. 1 in \cite{Akaho2021ApJ...909..210A}).}

\subsection{Subgrid Model for Fast Flavor Conversion}
\label{sec:subgrid}
As mentioned earlier, we take into account the effect of FFC with BGK subgrid modeling \cite{Bhatnagar1954PhRv...94..511B}, which is a relaxation-time approximation toward a certain asymptotic state.
The second term on the right hand side in Eq. \ref{eq_conservBoltz} is taken into account as follows
\begin{equation}
\label{eq:s_osc}
S_{\rm osc}\equiv -\frac{1}{\tau_\mathrm{as}}(f-f^\mathrm{as}),
\end{equation}
where $f^\mathrm{as}$ is the asymptotic distribution after FFC, and $\tau_\mathrm{as}$ is the timescale of FFC.
In this study, we compare three different prescriptions to take into account FFC effects as follows.

\subsubsection{BGK subgrid model with 4-species}
\label{sec:4spBGK}

We treat $\nu_e$, $\bar\nu_e$, $\nu_x$ and $\bar\nu_x$ distinctively, where $\nu_x=\nu_\mu=\nu_\tau$, $\bar\nu_x=\bar\nu_\mu=\bar\nu_\tau$ \footnote{BGK subgrid model with 6-species treatment was also suggested in \cite{Liu2025arXiv250318145L}}.
This model assumes that FFC occurs in a way to smear out ELN-XLN crossings \cite{Zaizen2023PhRvD.107j3022Z}.
The asymptotic states of the distribution functions $f^\mathrm{as}$ are written as
\begin{eqnarray}
\label{eq:asym1}
f_e^\mathrm{as} &=& \eta f_e+(1-\eta)f_x, \\
\label{eq:asym2}
\bar f_e^\mathrm{as} &=& \eta \bar f_e+(1-\eta)\bar f_x, \\
\label{eq:asym3}
f_x^\mathrm{as} &=& \frac{1-\eta}{2} f_e+\frac{1+\eta}{2}f_x, \\
\label{eq:asym4}
\bar f_x^\mathrm{as} &=& \frac{1-\eta}{2} \bar f_e+\frac{1+\eta}{2}\bar f_x,
\end{eqnarray}
where $\eta$ represents the survival probability, and the subscripts $e$ and $x$ represent electron-type and heavy-lepton-type neutrinos, respectively. The barred quantities represent those for antineutrinos.
In order to calculate $\eta$, we introduce following quantities 
\begin{eqnarray}
A&\equiv&\left|\frac{1}{8\pi^3}\int_{\Delta G<0} d(\cos\theta_\nu)d\phi_\nu \Delta G\right|, \\
B&\equiv&\frac{1}{8\pi^3}\int_{\Delta G>0} d(\cos\theta_\nu)d\phi_\nu \Delta G, \\
\end{eqnarray}
where $\Delta G$ is the ELN-XLN defined as
\begin{equation}
\Delta G\equiv \int(f_e-\bar f_e - f_x + \bar f_x)\epsilon^2d\epsilon.
\end{equation}
{As already mentioned, FFI is known to be equivalent to the existence of ELN-XLN crossings, i.e. both $A$ (negative $\Delta G$ part) and $B$ (positive $\Delta G$ part) take nonzero values. It is assumed that the flavor equipartition is achieved for the asymptotic state for the certain angles to make either of $A$ or $B$ become zero, and the distribution for the remaining angles are shifted to conserve the neutrino number. Whether $A$ or $B$ is eliminated, is determined depending on which part is larger.}

For the case with $B>A$ {(positive ELN-XLN density)},
\begin{equation}
\label{eq:BA1}
\eta = \left\{
\begin{array}{cc}
1/3 & (\Delta G<0), \\
1-2A/(3B) & (\Delta G\ge0),
\end{array}
\right.
\end{equation}
and for the case with $B<A$ {(negative ELN-XLN density)},
\begin{equation}
\label{eq:BA2}
\eta = \left\{
\begin{array}{cc}
1/3 & (\Delta G>0), \\
1-2B/(3A) & (\Delta G\le0).
\end{array}
\right.
\end{equation}
{
In general, $\eta$ depends on space, energy and momentum angles. In our prescription, it is determined from energy-integrated quantity and the energy dependence is dropped ($\eta=\eta(r,\theta_\nu)$ in 1D).}

The relaxation time is estimated as
\begin{equation}
\label{eq:tau}
\tau_\mathrm{as} \equiv \frac{2\pi}{\sqrt{AB}}.
\end{equation}
{This formula was motivated by two-beam model and has been used to estimate the growth rates ($2\pi/\tau_\mathrm{as}$) in the post-process analyses in the previous studies \cite{Nagakura2019ApJ...886..139N,Morinaga2020PhRvR...2a2046M,Harada2022ApJ...924..109H,Akaho2023ApJ...944...60A,Akaho2024PhRvD.109b3012A}.}
Hereafter, this model is called as \texttt{4spBGK}.

\subsubsection{BGK subgrid model with 3-species}
\label{sec:3spBGK}

In order to quantify the effect of the 3-species assumption, we also perform BGK subgrid model calculation by assuming $\nu_x=\bar\nu_x$.
The asymptotic distribution is determined by just imposing $f_{\nu_x}=f_{\bar\nu_x}$ in Eqs. \ref{eq:asym1}-\ref{eq:asym4}, which yields 
\begin{eqnarray}
\label{eq:asym3sp}
f_e^\mathrm{as} &=& \eta f_e+(1-\eta)f_x, \\
\bar f_e^\mathrm{as} &=& \eta \bar f_e+(1-\eta)f_x, \\
f_x^\mathrm{as} &=& \frac{1-\eta}{4} f_e+\frac{1-\eta}{4} \bar f_e+\frac{1+\eta}{2}f_x.
\end{eqnarray}
Note that this prescription assumes XLN$=0$, which leads to the lepton number violation.
The growth rate is estimated in the same way as the 4-species model, but since XLN is vanishing in 3-species case, the growth rate is determined only from ELN.
Hereafter, this model is called as \texttt{3spBGK}.

\subsubsection{Simple Flavor Equipartition with Density threshold}
\label{sec:3sprho11}

As mentioned earlier, simulations based on the approximate transport methods do not have sufficient information to detect FFI, and resort to a simple equipartition approach for region below a certain density threshold {\cite{Ehring2023PhRvD.107j3034E,Ehring2023PhRvL.131f1401E,Mori2025PASJ..tmp...18M,Qiu2025arXiv250311758Q}}.
We test the effects of such simplified approach on neutrino distribution and CCSN dynamics.

We follow the prescription proposed in \cite{Ehring2023PhRvD.107j3034E}.
First, number densities of $\nu_e$ ($n_{\nu_e}$) and $\bar\nu_e$ ($n_{\bar\nu_e}$) are compared. It is assumed that the subdominant species and $\nu_x$ reach the equipartition. 
In addition, the number density of the dominant species is shifted so that the ELN ($n_{\nu_e}-n_{\bar\nu_e}$) is conserved.
Thus, the lepton number is conserved unlike \texttt{3spBGK}.
This conservation of ELN makes FFI to {remain even after the occurrence of FFC}, which is different from the spirit of \texttt{3spBGK} and \texttt{4spBGK} that try to remove FFI.

By imitating the prescription, we determine the asymptotic states as follows.
For the case of $n_{\nu_e}>n_{\bar\nu_e}$ (equipartition for $\bar\nu_e$ and $\nu_x$),
\begin{eqnarray}
\label{eq:densitythreshold1a}
f^\mathrm{as}_{e}&=& f_e + \frac{2}{3}(f_x-\bar f_e), \\
\label{eq:densitythreshold1b}
\bar f^\mathrm{as}_{e}&=&\frac{\bar f_e+2f_x}{3},  \\
\label{eq:densitythreshold1c}
f^\mathrm{as}_{x}&=&\frac{\bar f_e+2f_x}{3}.
\end{eqnarray}
For the case of $n_{\nu_e}<n_{\bar\nu_e}$ (equipartition for $\nu_e$ and $\nu_x$) 
\begin{eqnarray}
\label{eq:densitythreshold2a}
f^\mathrm{as}_{e}&=&\frac{f_e+2f_x}{3}, \\
\label{eq:densitythreshold2b}
\bar f^\mathrm{as}_{e}&=&\bar f_e + \frac{2}{3}(f_x-f_e), \\
\label{eq:densitythreshold2c}
f^\mathrm{as}_{x}&=&\frac{f_e+2f_x}{3}.
\end{eqnarray}
{Above manipulation is performed for all energy and momentum angles uniformly.}
The prescription in \cite{Ehring2023PhRvD.107j3034E} modified fluxes in order to ensure the momentum conservation. 
However, the above prescription in Eqs. \ref{eq:densitythreshold1a}-\ref{eq:densitythreshold2c}
automatically satisfies the momentum conservation because we consider the flavor mixing between different species with the same momentum angles. This is the difference between Boltzmann and moment transport methods, where number density and the fluxes are independent variable for the latter.

The relaxation time of FFC cannot be estimated when the approximate transport methods are used. 
Hence instant conversion \cite{Ehring2023PhRvD.107j3034E,Ehring2023PhRvL.131f1401E,Mori2025PASJ..tmp...18M} or constant relaxation time \cite{Qiu2025arXiv250311758Q} is typically employed.
{Although we can self-consistently determine $\tau_\mathrm{as}$ (Eq. \ref{eq:tau}), we imitate previous studies by setting the constant time as} $\tau_\mathrm{as}=10^{-7}\,\mathrm{s}$. 

The density threshold is set to be $10^{11}\,\mathrm{g}\,\mathrm{cm}^{-3}$, because our model employed in this paper shows the appearance of FFI at $\rho\lesssim 10^{11}\,\mathrm{g}\,\mathrm{cm}^{-3}$ (see Fig. \ref{fig:orig-growth}). 
Hereafter, this model is called as \texttt{3sp$\rho$11}.


\subsection{Reference Model}
\label{sec:initial}
In this section, we explain how the initial CCSN profile was chosen and from what time the FFC simulations are performed.
Since it is known that the appearance of FFI is suppressed in spherically symmetric CCSN models
\footnote{{In the absence of convection, deleptonization is suppressed and $Y_e$ is kept higher. This situation makes the abundance of $\bar\nu_e$ to be smaller than $\nu_e$, and FFI is unlikely to appear in the post-shock region.}}, 
we pick up a snapshot from 2D CCSN simulations and angle-average it as the initial data for the 1D simulation.
We first run a 1D relaxation simulation (without FFC), and when appreciable FFI appears behind the shock wave, FFC subgrid models are started. {This is referred to as the fiducial model, hereafter.}
The details of the original 2D simulation are described in Sec. \ref{sec:initial2D}, and the results from 1D relaxation simulation is shown in Sec. \ref{sec:initial1D}.

\subsubsection{2D model}
\label{sec:initial2D}
We employ CCSN model of the progenitor with zero-age-main-sequence mass with $11.2\,\mathrm{M_\odot}$, taken from \cite{WoosleyRevModPhys.74.1015}.
The numerical details of 2D simulation is based on \cite{Nagakura2019ApJS..240...38N}, but with general relativistic gravity, and the detailed analysis of dynamics will be reported elsewhere.
The 2D model employs exactly the same neutrino-matter interactions and EOS, as the 1D simulations in this paper.
The mesh configurations in 2D are the same as in 1D for $r$, $\theta$ and $\epsilon$. In 2D, the zenith angle grid in the configuration space covers the range $\theta\in[0:\pi]$ with 128 mesh points, and the azimuth angle grid in momentum space covers the range $\phi_\nu\in[0:2\pi]$ with 6 mesh points.
We pick up a time snapshot of $t_\mathrm{pb}=270\,\mathrm{ms}$ after bounce.
The shock exists in $200\lesssim r\lesssim 500\,\mathrm{km}$ at this snapshot.

\subsubsection{1D Simulation from the angle-averaged 2D profile}
\label{sec:initial1D}

\begin{figure}
    \centering
    \includegraphics[width=\linewidth]{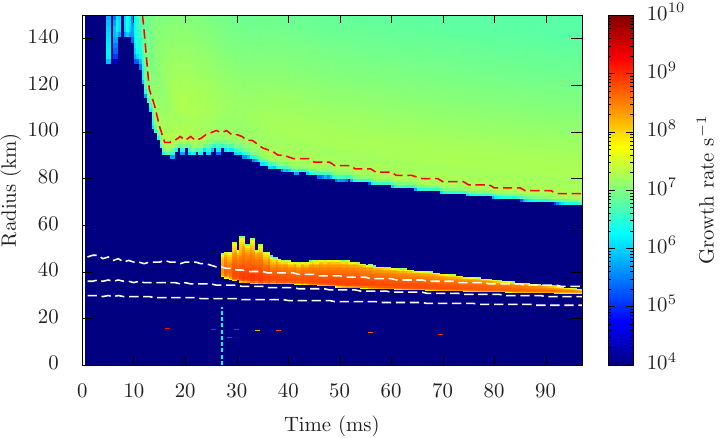}
    \caption{Time-radius map of the FFI growth rates for the 1D CCSN simulation without FFC. The time $t=0$ corresponds to the time 1D simulation started from the angle-averaged 2D profile at $t=270\,\mathrm{ms}$ after bounce. The red line represents the shock radius, and the white dashed lines correspond to the radius where the density is $\rho=10^{10}$, $10^{11}$, $10^{12}\,\mathrm{g}\,\mathrm{cm}^{-3}$. The vertical dashed line in cyan denotes the initial time where mixing simulation was performed.}
    \label{fig:orig-growth}
\end{figure}

Fig. \ref{fig:orig-growth} shows the time evolution of FFI growth rates ($2\pi/\tau_\mathrm{as}$) of 1D CCSN simulation without FFC. The time $t=0$ corresponds to the time 1D simulation started from the angle-averaged 2D profile, $t=270\,\mathrm{ms}$ after bounce.
Initially, the sudden disappearance of turbulence leads to rapid recession of shock wave, and it settles down at $t\sim15\,\mathrm{ms}$. 
We find that the FFI region appears at $t=27\,\mathrm{ms}$ around $r\sim40\,\mathrm{km}$.
This is caused by so-called type-II crossing \cite{Nagakura2021PhRvD.104h3025N}, where $\bar\nu_e$ dominates over $\nu_e$ for the radially outgoing direction and opposite for the ingoing direction. 
This type of FFI typically appears around the $\nu_e$ neutrino sphere, where $\bar\nu_e$ are half-decoupled from matter and become more forward-peaked than $\nu_e$. Appearance of this kind of FFI has already been reported in previous studies \cite{Nagakura2021PhRvD.104h3025N,Harada2022ApJ...924..109H,Akaho2023ApJ...944...60A,Akaho2024PhRvD.109b3012A}.
The FFI region outside the shock wave is generated by the type-I crossings where back-scatter of $\bar\nu_e$ makes it dominant over $\nu_e$ for the ingoing direction. 
{We particularly focus on this type-II FFI because the flavor composition of the outgoing neutrinos is changed and possibly affect the neutrino heating rates and the observed spectra.}
Note that the type-II FFC region shrinks with time faster than the PNS contraction, and disappears at $t\sim120\,\mathrm{ms}$.
This is because switching to 1D makes the equilibrium $Y_e$ value to be higher than the 2D case. 
We start the mixing simulation from the time right before type-II FFI appears at $\sim40\,\mathrm{km}$ (denoted as cyan vertical dashed line in Fig. \ref{fig:orig-growth}).

\begin{figure}
    \centering
    \includegraphics[width=\linewidth]{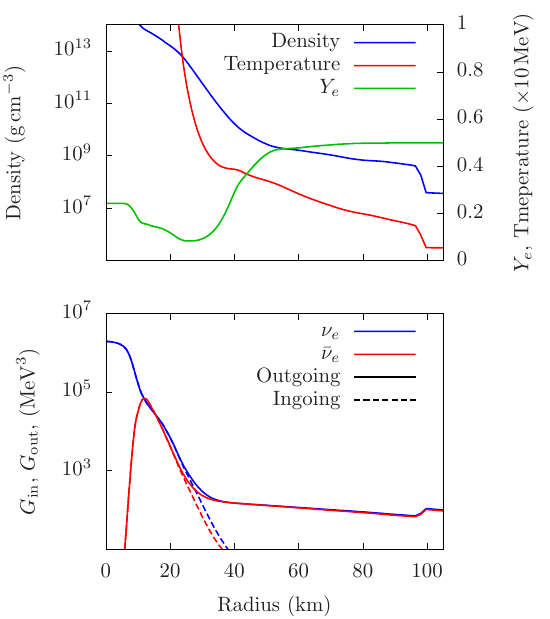}
    \caption{Radial profiles of the density, temperature, $Y_e$ (top) and the energy-integrated distribution functions (bottom), at the time of FFI appearance at $\sim40\,\mathrm{km}$ (denoted as cyan vertical line in Fig. \ref{fig:orig-growth}).}
    \label{fig:orig-rad}
\end{figure}

Hydrodynamics and neutrino profiles at that snapshot is presented in Fig. \ref{fig:orig-rad}. 
{
$G_\mathrm{out}$ and $G_\mathrm{in}$ are the energy-integrated distribution function for the radially outgoing and ingoing neutrinos, respectively, defined as
\begin{eqnarray}
&&G_\mathrm{out} \equiv \int f(\theta_\nu=0,\epsilon)\epsilon^2 d\epsilon. \\
&&G_\mathrm{in} \equiv \int f(\theta_\nu=\pi,\epsilon)\epsilon^2 d\epsilon.
\end{eqnarray}}
For the ingoing direction, $\nu_e$ dominates over $\bar\nu_e$. On the other hand, for the outgoing direction, $\bar\nu_e$ abundance slightly excess $\nu_e$ at $\sim40\,\mathrm{km}$, which generates FFI.

{Our fiducial model offers a more realistic scenario than previous studies that assumed a fixed hydrodynamical background. Such studies often suppose an initial deep crossing of neutrino flavors, a condition that is unachievable if FFC actually occurs. In contrast, our model solves the hydrodynamics simultaneously, allowing the background to evolve. This evolution gradually increases the abundance of $\bar\nu_e$ relative to $\nu_e$, which induces natural appearance of FFI.}

\subsubsection{Lower-$Y_e$ model}
As we shall show in Sec. \ref{sec:long}, the fiducial model does not show appreciable difference to the hydrodynamical profile between models with and without FFC.
Therefore, we additionally perform simulations with $Y_e$ artificially lowered by $10\%$.
Although this manipulation itself is artificial, this has clear physical motivation.
Multi-dimensional simulations with strong asymmetry sometimes show strong $\bar\nu_e$ emission over $\nu_e$, such as the lepton-number emission self-sustained asymmetry (LESA) \cite{Tamborra2014ApJ...792...96T}, or PNS kick \cite{Nagakura2019ApJ...880L..28N}.
Such asymmetry is known to generate FFI for a certain angle \cite{Nagakura2019ApJ...886..139N} and we test what occurs in such situations.

\section{Testing time integration methods}
The term corresponding to the flavor mixing $S_\mathrm{osc}$ is taken into account in the operator-splitting approach. 
After the time advancement of classical Boltzmann terms, namely advection and collision terms, the mixing term is added in BGK method. 
In this section, we test time integration methods and compare the results for different methods.
Sec. \ref{sec:timemethod} describes three approaches compared in this study (explicit, semi-implicit and implicit methods), and the results are compared in Sec. \ref{sec:timetest}.

\subsection{Time Advancement Procedure}
\label{sec:timemethod}

We use following notations. The symbol $f^\ast$ represents the temporal distribution function after the calculation of advection and collision terms. The symbol $f_{n+1}$ represents the distribution function after the flavor conversion, which is used for the advection and collision calculation in the next time step. {Time step width is represented with $\Delta t$.}

\subsubsection{Explicit method}

The simplest method is to estimate the FFC term in an explicit way;
\begin{equation}
f^{n+1} = f^\ast - \frac{\Delta t}{\tau_\mathrm{as}}(f^\ast-f^\mathrm{as}),
\end{equation}    
$f^\mathrm{as}$ is estimated from $f^\ast$, based on Eqs. \ref{eq:asym1}-\ref{eq:asym4}.
Hereafter, we refer to this time advancement method as the explicit method.

\subsubsection{Semi-implicit method}

Second method is the time advancement method proposed in \cite{Nagakura2024PhRvD.109h3013N}. 
{It showed reasonable agreement with the results with direct calculation of QKE.}
By estimating $f$ in the FFC term with $f^{n+1}$, discretized time evolution equation becomes
\begin{equation}
\frac{f^{n+1}-f^\ast}{\Delta t} = - \frac{f^{n+1}-f^\mathrm{as}}{\tau^\mathrm{as}},
\end{equation}
which yields the recurrence formula as 
\begin{equation}
f^{n+1} = \left(\frac{1}{\Delta t}+\frac{1}{\tau_\mathrm{as}}\right)^{-1}
\left(\frac{f^\ast}{\Delta t} + \frac{f^\mathrm{as}}{\tau_\mathrm{as}}\right).
\end{equation}  
We refer to this method as the semi-implicit because the asymptotic state is not evaluated using $f^{n+1}$.
As we shall show in Sec. \ref{sec:timetest}, this method is the most preferable among three methods.

\subsubsection{Implicit method}

If the asymptotic states are given as the linear function of distribution function, it is also possible to estimate the asymptotic states using $f^{n+1}$.
By re-writing Eqs. \ref{eq:asym1}-\ref{eq:asym4}, the time evolution equation becomes
\begin{equation}
\frac{1}{\Delta t}\left(
\begin{matrix}
f_e^{n+1}-f_e^\ast \\
f_x^{n+1}-f_x^\ast
\end{matrix}
\right) 
=
-\frac{1-\eta}{\tau_\mathrm{as}}
\left(
\begin{matrix}
1 & -1 \\
-\frac{1}{2} & \frac{1}{2}
\end{matrix}
\right) 
\left(
\begin{matrix}
f_e^{n+1} \\
f_x^{n+1}
\end{matrix}
\right) ,
\end{equation} 
which yields the recurrence formula
\begin{equation}
\left(
\begin{matrix}
f_e^{n+1} \\
f_x^{n+1}
\end{matrix}
\right)
 = 
\left(
\begin{matrix}
\tau^\mathrm{as}+\Delta t(1-\eta) & -\Delta t(1-\eta)\\
-\Delta t(1-\eta) & \tau^\mathrm{as}+\Delta t(1-\eta)
\end{matrix}
\right)^{-1}
\left(
\begin{matrix}
f_e^\ast \\
f_x^\ast
\end{matrix}
\right),
\end{equation}
Hereafter, we refer to this method as the implicit method.
It is usually considered that estimating more terms with $n+1$-th step leads to a more stable calculation.
However, in our case, this implicit method is not necessarily the most robust one unfortunately. This is because it does not satisfy the lepton number conservation unlike the semi-implicit method does. {This fact can be easily confirmed by calculating that $f^{n+1}_e-\bar f^{n+1}_e + 2 f^{n+1}_x - 2 \bar f^{n+1}_x$ differs from $f^\ast_e-\bar f^\ast_e + 2 f^\ast_x - 2 \bar f^\ast_x$.}

\subsection{Demonstration}
\label{sec:timetest}

\begin{figure}
    \centering
    \includegraphics[width=\linewidth]{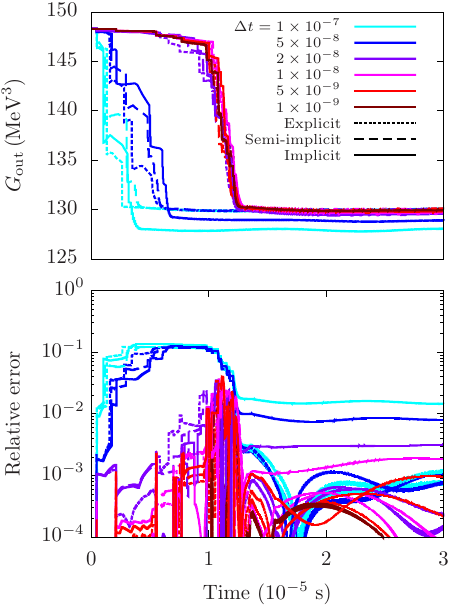}
    \caption{Time evolution of $G_\mathrm{out}$ at $r=41\,\mathrm{km}$ (top) and the relative error with respect to implicit model with $\Delta t=10^{-9}\,\mathrm{s}$ (Bottom). {Time $t=0$ corresponds to the start of the mixing simulation (vertical cyan line in Fig. \ref{fig:orig-growth}).} Colors represent the time step. The dotted, dashed, and solid lines represent results for explicit, semi-implicit, and implicit cases, respectively.}
    \label{fig:forward_comp}
\end{figure}

We compare three time advancement methods (Sec. \ref{sec:timemethod}), and also compare dependence of the results on time step width $\Delta t$.
The typical timescale employed in Boltzmann neutrino radiation hydrodynamics simulations is $\Delta t\sim10^{-8}-10^{-7}\,\mathrm{s}$, which can be longer than the timescale of FFC. 
This comparison is meant to determine how much $\Delta t$ is required.
We run short-time simulation from the fiducial model, which is the snapshot right before FFI appearance, as explained in Sec. \ref{sec:initial}.
The conversion model is \texttt{4spBGK}.
Fig. \ref{fig:forward_comp} shows the time evolution of $G_\mathrm{out}$.
{The value is for the radius $r=41\,\mathrm{km}$, which corresponds to the middle of type-II FFI region observed in Fig. \ref{fig:orig-growth}.}
The avalanche-like drop seen for all models corresponds to the appearance of ELN crossing and the conversion of $\nu_e$ into $\nu_x$, trying to smear out the crossing.
Models with $\Delta t =10^{-7}$ and $5\times10^{-8}\,\mathrm{s}$ shows earlier growth than models with smaller $\Delta t$. 
This behavior is natural because too large $\Delta t$ generates deeper ELN crossings, which makes the growth rates higher {($\tau_\mathrm{as}$ shorter)}, and the amount of mixed neutrinos is larger.
For models with $\Delta t\lesssim10^{-8}\,\mathrm{s}$, the balance between {the generation of FFI (by advection)} and {the elimination of FFI (by FFC subgrid model)} is well resolved, hence the results seems to be converged.
The explicit method tends to show relatively earlier avalanche.
{It is a known characteristic of explicit method to induce significant changes per time step in the presence of stiff terms, and with larger time steps, the FFC term indeed acts as a stiff term.}
The quasi-steady state is observed at $t\gtrsim2\times10^{-5}\,\mathrm{s}$, where the flavor mixing and the generation of FFI balances. This state is reasonably resolved in all models, except for the implicit models with $\Delta t =10^{-7}$, $5\times10^{-8}\,\mathrm{s}$.
This is because the lepton number is violated in the implicit method, as mentioned earlier.
Especially in our operator-splitting approach, the computational costs are exactly the same for three approaches.
The results suggests that the semi-implicit method is the most preferable.
\begin{figure}
    \centering
    \includegraphics[width=\linewidth]{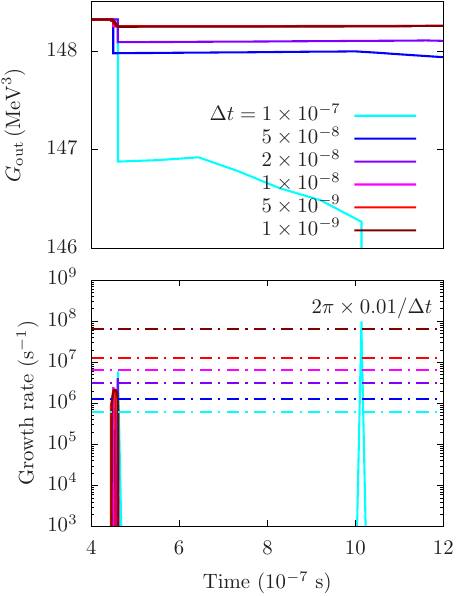}
    \caption{Top panel: a closer look of the top panel of Fig. \ref{fig:forward_comp}. Bottom panel: growth rates of FFC for the corresponding time range. The colors denote $\Delta t$, in the same way as Fig. \ref{fig:forward_comp}. Cases for implicit methods are only shown. As a reference, horizontal lines representing {$2\pi\times0.01/\Delta t$} are shown in the bottom panel.}
    \label{fig:forward_comp_fine}
\end{figure}

Detailed views of the time evolution of $G_\mathrm{out}$ and the growth rates at the onset of appearance of ELN crossing, are shown in Fig. \ref{fig:forward_comp_fine}.
As a reference, horizontal lines representing $2\pi\times 0.01/\Delta t$ are shown in comparison with the growth rates. 
If the growth rate is below the horizontal line for corresponding $\Delta t$, it means that FFC is resolved with at least $\sim100$ time steps. 
At the leftmost range $t=4\times10^{-7}\,\mathrm{s}$, all models show exactly the same values since the FFC has not occurred yet.
At $t\sim4.2\times10^{-7}\,\mathrm{s}$, growth rates of $\sim10^{6}\,\mathrm{s}^{-1}$ appear for all models.
Note that this FFI is just a temporal instability, which is soon smeared out by flavor conversion. 
At later time, crossings persistently appear and balance with flavor conversion.
Models with $\Delta t\geq 2\times10^{-8}$ show clear overshoot (over-conversion of $\nu_e$ into $\nu_x$) compared to models with smaller $\Delta t$.
The growth rates for those models overlap horizontal lines $0.01/\Delta t$, which suggests that $\Delta t \lesssim 0.01 \tau_\mathrm{as}$ is necessary to avoid the overshoot. 
However, as we have already seen, the quasi-steady state is well resolved with $\Delta t=10^{-7}\,\mathrm{s}$ for explicit or semi-implicit methods (Fig. \ref{fig:forward_comp}), so this strict time limitation seems to be necessary only for capturing the initial onset of FFC.
The simulations presented in subsequent sections are all performed with maximum $\Delta t=10^{-8}\,\mathrm{s}$ and semi-implicit time advancement method.

\section{Appearance of Instability and the Early-Time Flavor Evolution}
\label{sec:early}

In this section, we focus on the behavior in the early phase of FFC, for $t\lesssim5\times10^{-5}\,\mathrm{s}$.
The interplay of flavor mixing with hydrodynamics and the effects on CCSN dynamics based on longer simulations are discussed in Sec. \ref{sec:long}.
The fiducial model is chosen as the initial data in this section.
\begin{figure}
    \centering
    \includegraphics[width=\linewidth]{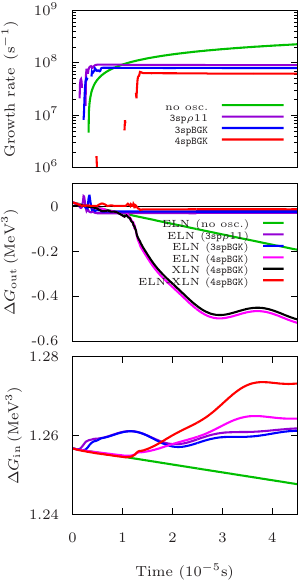}
    \caption{Time evolution of FFI growth rate (top) and $\Delta G_\mathrm{out}$ at $r=41\,\mathrm{km}$ (bottom). {The definition of the time is the same as Fig. \ref{fig:forward_comp}.} Note that only ELN are plotted for the no-oscillation model, \texttt{3sp$\rho$11} and \texttt{3spBGK} models {because XLN=0, by definition}. For \texttt{4spBGK}, ELN, XLN, ELN-XLN are plotted.}
    \label{fig:ELNXLN_timeevo}
\end{figure}
Time evolution of the FFI growth rates and $\Delta G_\mathrm{out}$ {(ELN-XLN for radially outgoing direction $\theta_\nu=0$)}, $\Delta G_\mathrm{in}$ {(ELN-XLN for radially ingoing direction $\theta_\nu=\pi$)} at $r=41\,\mathrm{km}$ are shown in Fig. \ref{fig:ELNXLN_timeevo}.
{Note that only ELN is plotted for 3-species models because XLN=0, by definition.}
Since $\Delta G_\mathrm{in}$ are kept positive in the time range shown here, the signature of $\Delta G_\mathrm{out}$ determines the existence of instability (as long as there is only one crossing).
{As for the type II FFI we are focusing now, ELN-XLN density is positive (Eq. \ref{eq:BA1}) and the radially outgoing neutrinos are mainly affected by the conversion.}
The no-oscillation model shows monotonically increasing FFI growth rate in the time range and the models with mixing treatments show convergence at a certain value.

Let us first focus on the first rise of the growth rate, at $t\lesssim10^{-5}\,\mathrm{s}$.
Models with 3-species assumption, \texttt{3sp$\rho$11} and \texttt{3spBGK} show earlier rise of the growth rate than \texttt{4spBGK}.
The growth rate is almost absent for \texttt{4spBGK} in $t\lesssim10^{-5}\,\mathrm{s}$ whereas FFI persistently exists for 3-species models.
This difference can be understood as follows. 
When the crossing is shallow, 4-species model can completely eliminate FFI by just distributing ELN to XLN. 
On the other hand, 3-species  subgrid models are not designed to erase FFI; \texttt{3sp$\rho$11} is designed to conserve ELN which does not change FFI growth rate, and \texttt{3spBGK} tries to decrease ELN but cannot completely erase it because of equalizing of $\nu_x$ and $\bar\nu_x$. 
This makes the growth rates to be overestimated, {which makes $\tau_\mathrm{as}$ too short, and results in the} over-conversion compared to 4-species case as we see later.
Note that the increase of growth rates for \texttt{3sp$\rho$11} and \texttt{3spBGK} are even earlier than the no oscillation model.
This is because the global radiation field is already different; the FFC occurring at inner radius makes $\nu_e$ and $\bar\nu_e$ abundance closer, which makes a preferable condition for FFI {than the no-oscillation model}.

Let us now focus on the phase at $t\gtrsim10^{-5}\,\mathrm{s}$, where the growth rates are converged for all mixing models.
As can be seen in the middle panel, ELNs for \texttt{3sp$\rho$11} and \texttt{3spBGK} are converged as well as the growth rates. 
On the other hand, the converged value for \texttt{4spBGK} is ELN-XLN {(red line)}, and ELN {(magenta line)} and XLN {(black line)} both evolve in time. 
This clearly suggests that 3-species assumption results in qualitatively different flavor evolution compared to 4-species case.

\begin{figure*}
    \centering
    \includegraphics[width=\linewidth]{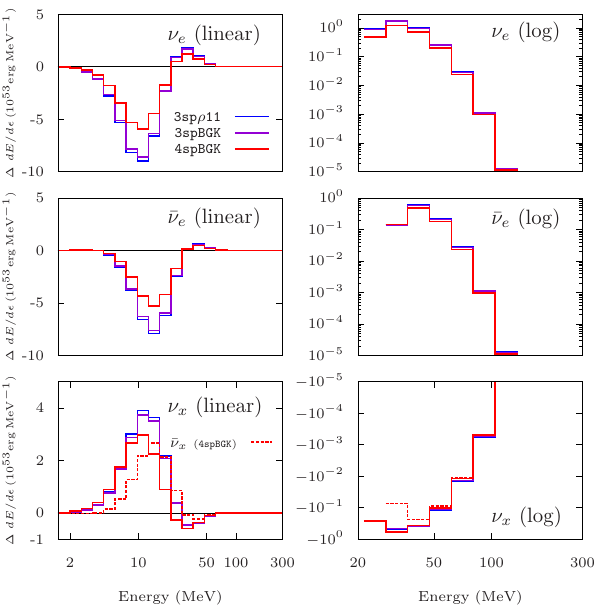}
    \caption{Difference of energy spectrum {(total neutrino energy per energy bin)} with respect to the no-oscillation model at $2\,\mathrm{ms}$ from the starting time of the mixing simulation. $\nu_e$, $\bar\nu_e$, $\nu_x$ are shown in top, middle, bottom panels, respectively. Vertical scales are linear for left panels, and logarithmic for right panels in order to focus on high energy region. Different colors indicate different mixing schemes. The result for $\bar\nu_x$ is shown with dashed lines in the bottom panels only for \texttt{4spBGK}.}
    \label{fig:enespec_short}
\end{figure*}

\begin{figure}
    \centering
    \includegraphics[width=\linewidth]{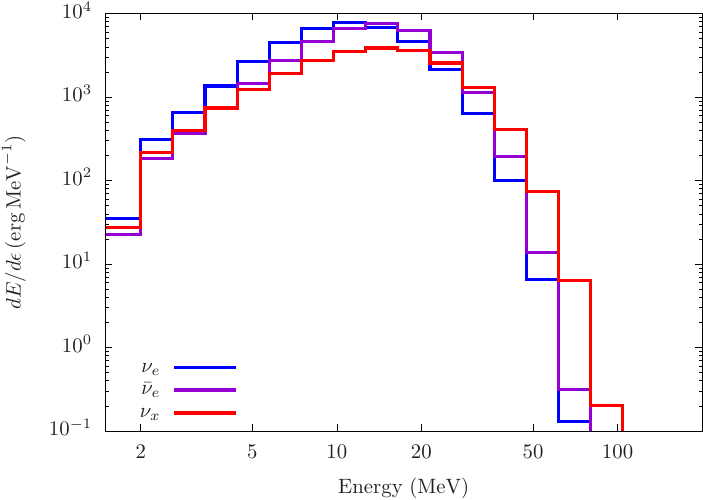}
    \caption{{Energy spectrum of the no-oscillation model at $2\,\mathrm{ms}$ from the starting time of the mixing simulation, which was used for the reference in Fig. \ref{fig:enespec_short}.}}
    \label{fig:enespec_orig}
\end{figure}

In Fig. \ref{fig:enespec_short}, we plot the difference of the energy spectrum {(total neutrino energy per energy bin)} with respect to the no-oscillation model. 
{For reference, we also show the energy spectra for the no-oscillation model in Fig. \ref{fig:enespec_orig}. It is clear that $\nu_x$ has harder energy spectrum compared to $\nu_e$ and $\bar\nu_e$.}
All mixing models show the same trend; $\nu_e$ and $\bar\nu_e$ with energy $\epsilon\sim10\,\mathrm{MeV}$ is reduced and those with energy $\epsilon\gtrsim30\,\mathrm{MeV}$ is enhanced. 
The enhancement/reduction trend is opposite for $\nu_x$ ($\bar\nu_x$) because the spectrum change is caused by the mixing between $\nu_e$ ($\bar\nu_e$) and $\nu_x$ ($\bar\nu_x$). 
Different subgrid models show quantitative differences. 
Although the differences between \texttt{3sp$\rho$11} and \texttt{3spBGK} is minor, they show clear deviation from \texttt{4spBGK}.
This is because, the 3-species assumption causes over-conversion because the growth rate is overestimated, as we discussed previously.

The spectrum of $\nu_x$ and $\bar\nu_x$ show clear difference in \texttt{4spBGK}; the enhancement peak is shifted to the higher energy side for $\bar\nu_x$.
This is natural because $\bar\nu_x$ is mixed with $\bar\nu_e$, which has higher average energy than $\nu_e$.
This suggests that $\nu_x$ and $\bar\nu_x$ should be separately treated.

\section{Late phase analysis}
\label{sec:long}
In this section, we show results of longer time simulations ($\sim100\mathrm{ms}$) and investigate the time evolution of growth rates (Sec. \ref{sec:growthrateevo})), effects of FFC on fluid dynamics (Sec. \ref{sec:fluid}) and the neutrino emission properties (Sec. \ref{sec:neutrinoemission}).
We discuss the results for not only fiducial model but also lower-$Y_e$ one in this analysis.

\subsection{Evolution of growth rates}
\label{sec:growthrateevo}
\begin{figure*}
    \centering
    \includegraphics[width=\linewidth]{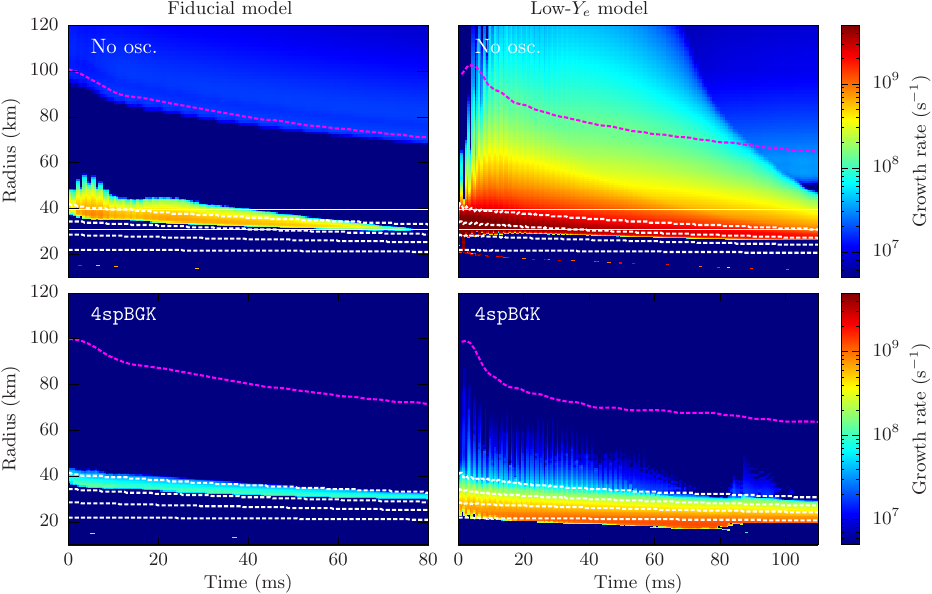}
    \caption{Time-radius map of the growth rates for the fiducial (left) and lower-$Y_e$ model (right). {The definition of the time is the same as Fig. \ref{fig:forward_comp}.} Red dashed lines represent the shock radii, and the white dashed lines represent the radius corresponding to the density of $10^{10}$, $10^{11}$, $10^{12}$, $10^{13}\,\mathrm{g}\,\mathrm{cm}^{-3}$.}
    \label{fig:growth_long}
\end{figure*}

Fig. \ref{fig:growth_long} shows the time-radius maps of the FFI growth rates for the fiducial model and the lower-$Y_e$ model. 
For both models, the FFI growth rates are lower with FFC subgrid modeling and kept at a certain value, balanced with advection terms which try to enhance FFI growth rates.
The noticeable difference in the lower-$Y_e$ model is that the FFI region is extended deeper in the core compared to the no-oscillation model.
In the no-oscillation model, FFI is observed at $\rho\lesssim10^{12}\,\mathrm{g}\,\mathrm{cm}^{-3}$ and $\rho>10^{13}\,\mathrm{g}\,\mathrm{cm}^{-3}$, where FFI is absent in the intermediate region ($10^{12}\lesssim\rho\lesssim10^{13}\,\mathrm{g}\,\mathrm{cm}^{-3}$).
With FFC subgrid modeling, on the other hand, FFC in the inner region affects radially outgoing propagating neutrinos and invoke FFI in the intermediate region.
This FFC greatly facilitates neutrino cooling for the lower-$Y_e$ model, as we see later.

\subsection{Effects onto fluid dynamics}
\label{sec:fluid}
\begin{figure*}
    \centering
    \includegraphics[width=\linewidth]{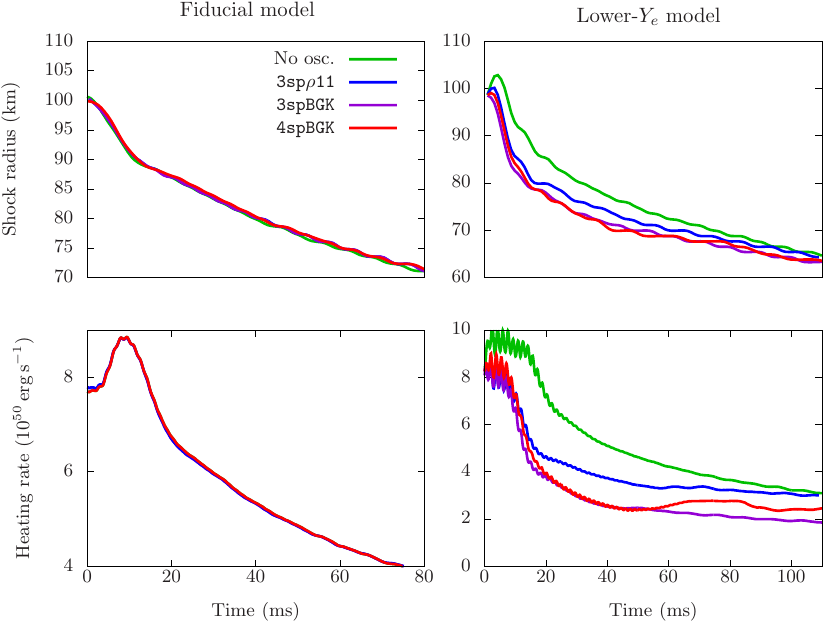}
    \caption{Time evolution of the shock radius (top) and neutrino heating rate (bottom) for the fiducial model (left) and lower-$Y_e$ model (right). {The definition of the time is same as Fig. \ref{fig:forward_comp}.}}
    \label{fig:shockevo}
\end{figure*}

Time evolution of the shock radius and the total neutrino heating rate {(absorptivity-emissivity integrated over the gain region)} are shown in Fig. \ref{fig:shockevo}. 
The fiducial model does not show visible difference whereas the mixing simulations in lower-$Y_e$ model shows smaller shock radii and lower neutrino heating rates.
Note that the oscillatory behavior in the lower-$Y_e$ model is caused by the PNS oscillation due to the artificial $Y_e$ profile. 
It is an artifact of our initial condition but does not alter our discussion qualitatively.

\begin{figure*}
    \centering
    \includegraphics[width=\linewidth]{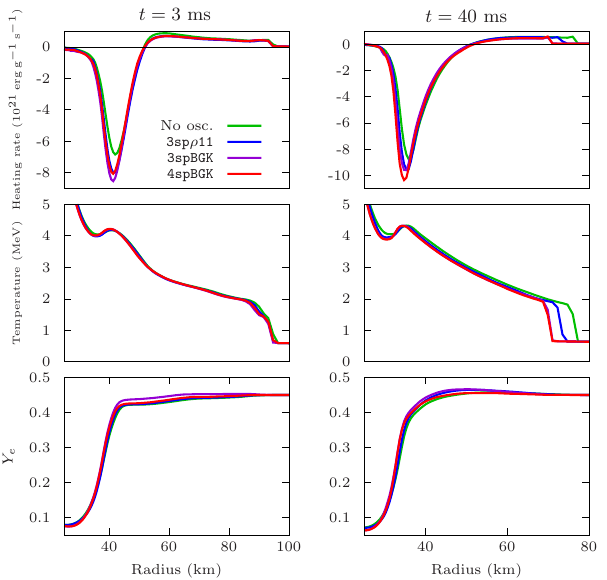}
    \caption{Radial profiles of neutrino heating rate {per unit mass} (top), temperature (middle), $Y_e$ (bottom) for the $t=3\,\mathrm{ms}$ (left) and $t=40\,\mathrm{ms}$ (right) after the mixing simulation, for the lower-$Y_e$ model. The time indicates the time from the start of the mixing simulation.}
    \label{fig:heat_rad}
\end{figure*}

For the lower-$Y_e$ model, radial profiles of the neutrino heating rates {per unit mass} are shown in Fig. \ref{fig:heat_rad} for two time snapshots $t=3$ and $40\,\mathrm{ms}$, after the start of the mixing simulation. 
The snapshot at $t=3\,\mathrm{ms}$ is supposed to illustrate the profile before the FFC alters the hydrodynamical profile, and $t=40\,\mathrm{ms}$ shows the profile reflecting FFC effects.
{The region with a positive neutrino heating rate is the gain region, while the region with a negative rate is where neutrino cooling is effective.}
At $t=3\,\mathrm{ms}$, all mixing simulations show deeper dip of the cooling rate than the no-oscillation model, and the heating rate is reduced.
The conversion of $\nu_e$ ($\bar\nu_e$) into $\nu_x$ ($\bar\nu_x$) facilitates easier decoupling for the former, which results in the enhancement of the cooling rate. 
This feature has already been reported previously in the simulations under the fixed hydrodynamical background \cite{Nagakura2023PhRvL.130u1401N,Xiong2024PhRvD.109l3008X}.
Lower heating rate leads to shrinkage of the shock radius and contraction of the gain region, which can be seen at $t=40\,\mathrm{ms}$.
Although the discrepancies in the density profiles between the methods are still minor at the $t=40\,\mathrm{ms}$ snapshot, they are anticipated to become more pronounced when the simulation is continued to a duration of several $100\,\mathrm{ms}$.

{It is worth mentioning that it is premature to regard the feature shown in Fig. \ref{fig:shockevo} (FFC has little effects in the fiducial model and negative effects onto the shock evolution in lower-$Y_e$ case) as the universal characteristics. 
Unveiling the roles of FFC onto CCSNe requires multi-dimensional simulations, which will be reported in the near future.}

\subsection{Neutrino emission properties}
\label{sec:neutrinoemission}
\begin{figure*}
    \centering
    \includegraphics[width=\linewidth]{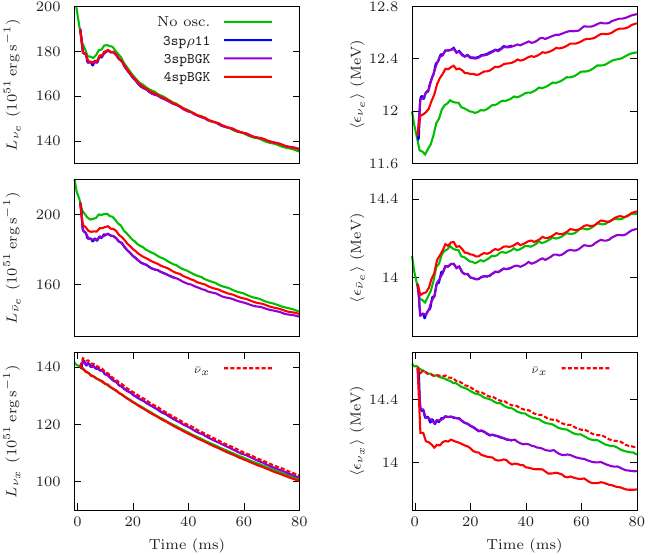}
    \caption{Time evolution of emitted neutrino luminosities {$L_\nu$} (left) and mean energies {$\langle\epsilon_\nu\rangle$} (right) for $\nu_e$ (top), $\bar\nu_e$ (middle), $\nu_x$ (bottom). Quantities of $\bar\nu_x$ for \texttt{4spBGK} are shown in the bottom panels. {The definition of the time is same as Fig. \ref{fig:forward_comp}.}}
    \label{fig:lumi_fid}
\end{figure*}

\begin{figure*}
    \centering
    \includegraphics[width=\linewidth]{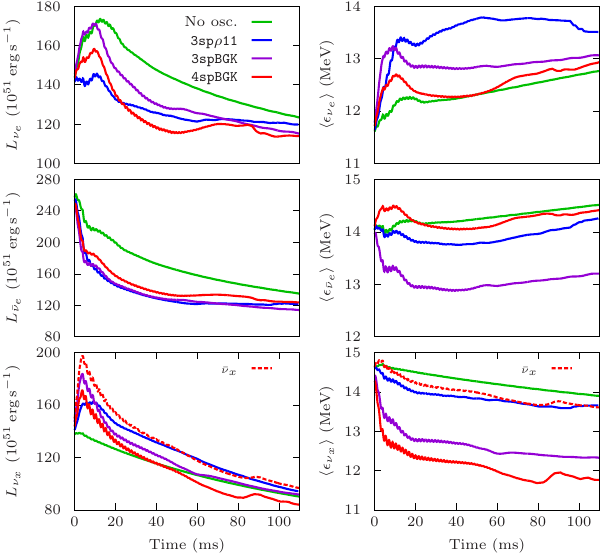}
    \caption{Same as Fig. \ref{fig:lumi_fid} but for the lower-$Y_e$ model.}
    \label{fig:lumi_low}
\end{figure*}

Emitted neutrino luminosities and mean energies are shown in Figs. \ref{fig:lumi_fid} and \ref{fig:lumi_low}, for the fiducial and lower-$Y_e$ models, respectively. 
{They are estimated at radius $r=500\,\mathrm{km}$.}
The properties of the emitted neutrinos show clear difference when FFC is considered, both for the fiducial and lower-$Y_e$ models. 

Let us first focus on the fiducial model (Fig. \ref{fig:lumi_fid}).
Luminosity of $\nu_e$ does not show large difference whereas $\bar\nu_e$ luminosity is lower for all mixing simulations, and $\nu_x$ (3-species case), $\bar\nu_x$ (4-species case) exhibits the enhancement. 
As for the average energy, $\nu_e$ shows a clear enhancement and $\nu_x$ shows a clear reduction. 

Both $\bar\nu_e$ and $\bar\nu_x$ mean energies are slightly enhanced in \texttt{4spBGK}, which is unlikely to be realized if the hydrodynamical profile is fixed.
This feature is due to faster contraction of PNS facilliated by FFC, which makes the matter temperature slightly higher.
Note that this slight enhancement of the mean energy is smeared out in the 3-species case because the mixing with $\nu_e$, which has much lower average energy, occurs.
{As mentioned earlier, the effect of FFC on the PNS profile is anticipated to be more pronounced if FFC persists for a longer duration. In that case, the properties of the emitted neutrinos are also likely to show larger differences with and without FFC.}

The luminosities and average energies for \texttt{3sp$\rho$11} and \texttt{3spBGK} almost coincide for all species.
This can be understood as follows.
In the type-II FFC region we are discussing now, ELN-XLN density is positive (Eq. \ref{eq:BA1}) and $\Delta G$ is negative for the outgoing direction.
Thus, the BGK model assumes simple equipartition (upper case in Eq. \ref{eq:BA1}), which results in almost same prescription as \texttt{3sp$\rho$11} for the outgoing direction.
However, as we see later in the lower-$Y_e$ model, this agreement breaks down when FFC alters the hydrodynamical profile itself because the difference of the ingoing neutrinos become important.

We now discuss the neutrino emission properties for the lower-$Y_e$ model (Fig. \ref{fig:lumi_low}).
The hierarchy of the enhancement or reduction of luminosities is more complicated than the fiducial model (Fig. \ref{fig:lumi_fid}).
This is because, in the fiducial model, the FFC causes the conversion of neutrinos mainly for free-streaming neutrinos. On the other hand, in the lower-$Y_e$ model, FFC changes the hydrodynamical profile {because neutrino decoupling is modified by FFC in the deeper core (Fig. \ref{fig:growth_long})}.
The trend is still similar to the fiducial model; conversion of $\nu_e$ and $\bar\nu_e$ into $\nu_x$ (and $\bar\nu_x$) causes the decrease of luminosities for the former and the enhancement of the latter. 
The average energies tend to be higher for the former and lower for the latter.

The degree of enhancement/reduction of mean energy is a few MeV, which is much higher than the fiducial model.
Another notable difference from the fiducial model is that \texttt{3sp$\rho$11} and \texttt{3spBGK} show clear deviation. 
The equipartition assumption in \texttt{3sp$\rho$11} can only properly treat conversion of outgoing neutrinos, and the difference in the ingoing direction alters the hydrodynamical profile, which results in the neutrino emission properties.
This indicates that angle-dependent subgrid modeling is necessary.

\section{Summary and Discussion}
\label{sec:summary}

We performed Boltzmann neutrino radiation hydrodynamics simulations of CCSN with the subgrid modeling of FFC in spherical symmetry. 
We first tested the effects of time discretization on treating FFC.
For the model tested, we found that $\Delta t\lesssim2\times10^{-8}\,\mathrm{s}$ is required to correctly capture the first appearance of FFI. 
The explicit method tends to overestimate the conversion compared to other two methods, and the implicit method failed to capture quasi-steady distribution for relatively large $\Delta t$. We conclude that the semi-implicit method is the most preferable.

Next, we compared the behavior of different subgrid modeling prescriptions for the early phase.
We found that the 3-species models, which aims to erase ELN crossings, behave differently from the 4-species models where ELN-XLN governs the instability.
This makes the growth rates rise earlier for 3-species models, and also the growth rate for the quasi-steady state is kept higher.

Finally, we investigated the effect of FFC on CCSN dynamics and neutrino emission properties, for the fiducial model and the lower-$Y_e$ model (with $Y_e$ lowered by $10\,\%$).
For the fiducial model, the shock radii and the heating rate did not show clear difference. 
In the lower-$Y_e$ model, on the other hand, all mixing simulations showed lower neutrino heating rates and hence smaller shock radii. 
The neutrino luminosities and neutrino mean energies showed difference for both models.
Overall trend is that the 3-species assumption tends to overestimate the conversion of $\nu_e$ and $\bar\nu_e$. In addition, $\nu_x$ and $\bar\nu_x$ shows clearly different luminosities and mean energies, implying that 4-species assumption should be used to accurately capture FFC effects.

We close this paper by noting some limitations. 
First, our calculations were limited to spherical symmetry, since this was meant to systematically compare different numerical prescriptions. 
However, the occurrence of FFC and the nature of CCSN dynamics is multi-dimensional.
We will perform multi-dimensional CCSN simulation with BGK subgrid modeling in the near future.

\begin{figure}
    \centering
    \includegraphics[width=\linewidth]{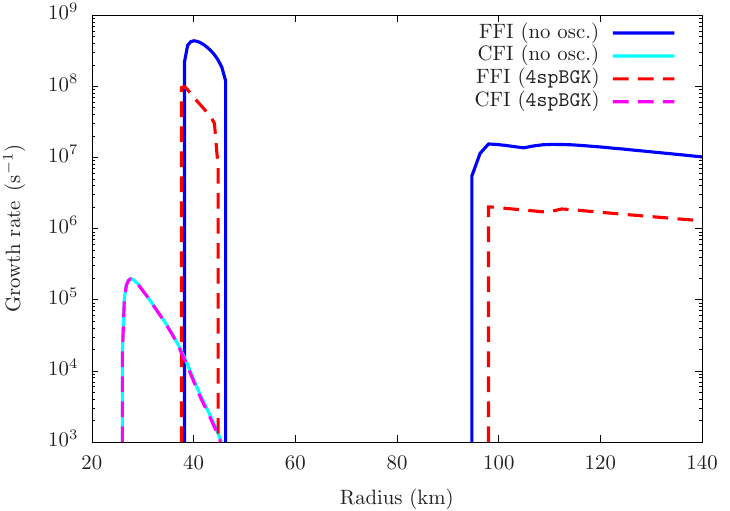}
    \caption{Radial profile of growth rates of FFI and CFI, $2\,\mathrm{ms}$ after the mixing simulation, for the fiducial model.}
    \label{fig:FFICFI}
\end{figure}

Second, the effect of the collisional flavor instability (CFI) \cite{Johns2023PhRvL.130s1001J} should be also taken into account.
For reference, we compare FFI and CFI growth rates in Fig. \ref{fig:FFICFI} between models with and without FFC for the fiducial model.
The linear growth rates of CFI was estimated by a formulae proposed in \cite{Liu2023PhRvD.107l3011L}, which was also used in the analyses in \cite{Liu2023PhRvD.108l3024L,Akaho2024PhRvD.109b3012A,Liu2024PhRvD.110d3039L}.
FFI growth rate is lower for the \texttt{4spBGK} model than the no-oscillation one, whereas CFI growth rate remains almost the same. This is because the CFI growth rate is weakly dependent on the angular distribution.
Although the CFI growth rate is smaller than that of FFI even after FFC, we cannot judge its importance solely from the growth rate.
The flavor conversion with the existence of CFI is expected to make a variety of asymptotic states \cite{Kato2024PhRvD.109j3009K,Zaizen2025PhRvD.111j3029Z,Froustey2025arXiv250516961F}.
We are planning to incorporate CFI effects on the subgrid model and investigate its effects of CCSNe in the near future (see also \cite{Wang2025arXiv250701100W}).

\begin{acknowledgments}
We thank Masamichi Zaizen for fruitful discussions.
Results of the Boltzmann radiation hydrodynamics simulations in this paper would not be available without Wakana Iwakami, Akira Harada, Shun Furusawa, Hirotada Okawa, Hideo Matsufuru and Kohsuke Sumiyoshi.
This work used high performance computing resources provided by Fugaku supercomputer at RIKEN, the Wisteria provided by JCAHPC through the HPCI System Research Project (Project ID: 240041, 240079, 240219, 240264, 250006, 250166, 250191, 250226, JPMXP1020200109, JPMXP1020230406), the FX1000 provided by Nagoya University, Cray XC50 and XD2000 at the National Astronomical Observatory of Japan (NAOJ), the Computing Research Center at the High Energy Accelerator Research Organization (KEK), Japan Lattice Data Grid (JLDG) on Science Information Network (SINET) of National Institute of Informatics (NII), Yukawa Institute of Theoretical Physics.
This work is supported by JSPS KAKENHI Grant No. JP24K00632.
H. N. is supported by Grant-in-Aid for Scientific Research (23K03468).
S. Y. is supported by Grant-in-Aid for Scientific Research (25K01006).
\end{acknowledgments}




\bibliography{apssamp}

\end{document}